\documentclass[preprint,12pt]{elsarticle} % elsarticle
\usepackage{float}
\usepackage{graphicx}
\usepackage{amssymb}
\usepackage{hyperref}
\usepackage{longtable}
\usepackage{pdflscape}
\usepackage{multirow, booktabs}
\usepackage{multicol}

\usepackage{tikz} % for tikz
\usepackage{graphicx}
\usetikzlibrary{positioning}

\journal{Journal of Finance}

\begin{document}

\begin{frontmatter}


\title{Handling missing data in Burundian sovereign bond market}

%% \tnotetext[label1]{}\author{Name\corref{cor1}\fnref{label2}}


\author[1,2]{Irene Irakoze\corref{mycorrespondingauthor}}
\cortext[mycorrespondingauthor]{Corresponding author}
\ead{irene.irakoze@ub.edu.bi}
\author[3]{Rédempteur Ntawiratsa}
\author[2,4]{David Niyukuri}

\address[1]{Actuarial Department, Institute of Applied Statistics, University of Burundi, Bubanza, Burundi }
\address[2]{Doctoral School, University of Burundi, Bujumbura, Burundi}
\address[3]{Department of Economics, Faculty of Economics and Management, University of Burundi, Bujumbura}
\address[4]{Department of Mathematics, Faculty of Science, University of Burundi, Bujumbura, Burundi }

% \address[5]{The South African Department of Science and Technology-National Research Foundation (DST-NRF) Centre of Excellence in Epidemiological Modelling and Analysis (SACEMA), Stellenbosch University, Cape Town, South Africa}

\begin{abstract}

Constructing an accurate yield curve is essential for evaluating financial instruments and analyzing market trends in the bond market. However, in the case of the Burundian  sovereign bond market, the presence of missing data poses a significant challenge to accurately constructing the yield curve. In this paper, we explore the limitations and data availability constraints specific to the Burundian sovereign market and propose robust methodologies to effectively handle missing data. The results indicate that the Linear Regression method, and the Previous value method perform consistently well across variables, approximating a normal distribution for the error values. The non parametric Missing Value Imputation using Random Forest (missForest) method performs well for coupon rates but poorly for bond prices, and the Next value method shows mixed results. Ultimately, the Linear Regression (LR)  method is recommended for imputing missing data due to its ability to approximate normality and predictive capabilities. However, filling missing values with previous values has high accuracy, thus, it will be the best choice when we have less information to be able to increase accuracy for LR. This research contributes to the development of financial products, trading strategies, and overall market development in Burundi by improving our understanding of the yield curve dynamics.

% The proposed solution has the potential to enhance transparency, reliability, and efficiency in the Burundian  bond market by providing more accurate yield curve estimations, aiding investors, market participants, and policymakers in making informed decisions. Additionally,

% Limitation de formules pour calculer le pri, redement, coupon

% Besoin d'avoir de donnee complete i on veut utilier Bootrap method pour calculer zero-coupon

% 

\end{abstract}


\end{frontmatter}

% \linenumbers

\section{Introduction}

For more than two decades, Burundi, through its central bank, the Bank of the Republic of Burundi (BRB), has been issuing treasury securities (\cite{instr2017, conv2018}). It started with short-term securities called treasury bills, with a maturity of 13, 26 and 52 weeks. Issuance has been generally regular, every Wednesday of the week. Demand was mainly from commercial banks. For a little less than a decade, the maturity has been progressively extended, even to the current 10 years. Note that when the maturity exceeds one year, the securities are called bonds. The timing of bond issuance is identical to that of treasury bills, but the range of maturities is far from complete. In countries with tight and illiquid financial market, the issuance is not regular. Besides that, when issuing treasury securities, some maturities are absent from the results of the auctions either because they are evaded from the offer, or because the prices proposed by the issuer do not meet with the approval of the demand. The result is an unbalanced database in terms of maturity. This imbalance is a major handicap when it comes to calculating the yields of zero-coupon bonds, which are a prerequisite for the construction of the yield curve \cite{voloshyn2015direct}.

The construction of an accurate yield curve is of paramount importance in evaluating financial instruments and analyzing market trends, particularly in the context of the bond market. However, the presence of missing data in the Burundian sovereign bond market poses a significant challenge to constructing a reliable yield curve. This paper aims to address this issue by exploring methodologies to effectively handle missing data and construct a robust yield curve for the Burundian sovereign bond market.

The Burundian sovereign bond market plays a vital role in the country's economy, providing a platform for capital mobilization, investment, and risk management. However, the unique dynamics and challenges of this emerging market present distinct obstacles to yield curve construction. Inadequate data availability and limitations in data collection practices contribute to missing data, which in turn hinders accurate yield curve estimations.

The objective of this research is to propose robust methodologies that mitigate the impact of missing data on yield curve estimations in the Burundian sovereign bond market. By employing suitable statistical techniques, we aim to enhance the transparency, reliability, and efficiency of the market by providing more accurate yield curve estimations. This, in turn, will aid investors, market participants, and policymakers in making informed decisions regarding financial instruments, investment strategies, and risk management.

%The limitations and data availability constraints specific to the Burundian financial bond market will be thoroughly examined in this study. Understanding the intricacies of data collection and availability is crucial for addressing missing data effectively. By exploring these factors, we can identify the gaps and challenges associated with data collection practices in Burundi, contributing to the overall improvement of the data infrastructure in the country's financial market.

To achieve our objective, we employ a range of statistical techniques tailored to the characteristics and limitations of the Burundian sovereign bond market. These methodologies will enable us to handle missing data effectively and construct a reliable yield curve. % By providing accurate yield curve estimations, this research will facilitate the evaluation of financial products, the development of trading strategies, and the overall advancement of the Burundian financial bond market.

% Moreover, this study has broader implications for the development of the financial sector in Burundi. By improving our understanding of the yield curve dynamics, we contribute to the enhancement of financial products, the formulation of effective trading strategies, and the overall market development. The insights gained from this research will aid policymakers in designing appropriate regulations and policies that foster a sustainable and inclusive financial system in Burundi.

Ultimately, this paper addresses the issue of handling missing data in the context of constructing a reliable yield curve for the Burundian financial bond market. By proposing robust methodologies and employing suitable statistical techniques, we aim to mitigate the impact of missing data and provide more accurate yield curve estimations. The outcomes of this research have the potential to enhance transparency, reliability, and efficiency in the Burundian sovereign bond market, facilitating informed decision-making for investors, market participants, and policymakers, while contributing to the overall development of the financial sector in Burundi.

%The purpose of our work is explore best approaches or techniques to fill in the missing data, namely prices, coupon rates and bond yields. To do this, we have tried several methods that are abundant in the scientific literature, and are explained in the methodology section. In the following subsection, we discuss the characteristics of Burundian financial market, and the need for a building a data set of treasury securities with full maturities.

% The rest of this article is organized as follows: a review of the literature on the interpolation methods used in the construction of the interest rate curve, the presentation of the Burundian treasury securities market and the related database, the results of the estimation of missing data and the conclusion.

%\subsection{Problem statement}

% The lack of a benchmark yield curve that can provide price signals to potential corporate bond issuers hinders the development of the corporate bond market \cite{rojas2014towards}. 

%4 Besoin d'avoir de donnee complete i on veut utilier Bootrap method pour calculer zero-coupon
The  bootstrapping technique has been used to construct the zero coupon yield curve that reflect the actual daily yield movements in the bond market \cite{hagan2008methods}. However, for this to be possible, we must have a complete series of  zero coupon rates, prices of bonds and the corresponding maturities what is not the case for the Burundian bond market due to its illiquid nature.

% 7 Besoin de develolopper les emthods pour faire face aux donnees manquqntes ppour calculer le tau zero coupon

% Financial markets are highly complex, and their performance is determined not only by how their institutions perform their functions, but also by various domestic macroeconomic, political, and other factors. An illiquid market is characterized by assets that cannot be traded indefinitely. Low liquidity markets lead to increased loss aversion due to low trading frequency. An illiquid market limits arbitrage in normal times as investors lose confidence in the market because an illiquid asset can be initially tradable, becomes untradable again after a fixed horizon or not at all after a set period of time. However, due to the complex, illiquid and incomplete nature of the emerging financial market, there is an even greater need to develop models based on imputing missing data to calculate the zero coupon rate.  Thus, the purpose of this work is to develop a model that will be used to impute missing data for the Burundian treasury bill and bond market.

The construction of an accurate yield curve is crucial for analyzing and valuing financial instruments in the bond market. The Burundian sovereign bond market, data availability is limited while the presence of missing data poses a significant challenge in accurately constructing the yield curve. This work aims to address the issues related to missing data and its impact on constructing a reliable yield curve for the Burundian sovereign bond market.

The Burundian  bond market serves as a vital platform for issuing and trading fixed-income securities, providing an avenue for raising capital for both government and corporate entities. However, due to factors such as data collection limitations, reporting delays, incomplete data records, the availability of historical bond market data in Burundi is insufficient and contains gaps. Missing data can occur in various forms, including absent observations for specific time periods, incomplete information on bond characteristics, or insufficient transaction data.

The presence of missing data poses challenges when attempting to construct a yield curve, which is a graphical representation of the relationship between bond yields and their respective maturities. An accurate yield curve is crucial for estimating market interest rates, valuing bonds, and conducting risk assessments \cite{evans2007economic}. However, if missing data is not appropriately handled, it can lead to biased or inaccurate yield curve estimations, potentially impacting investment decisions, risk management strategies, and overall market efficiency. 

The problem at hand requires the development of robust methodologies and techniques to handle missing data effectively in the construction of the yield curve for the Burundian financial bond market. The proposed solution considers the unique characteristics of the market, data availability constraints, and statistical techniques suitable for addressing missing data issues.

Addressing this problem contributes to enhancing the transparency, reliability, and efficiency of the Burundian bond market. Providing accurate yield curve estimations despite the presence of missing data allows market participants, investors, and policymakers to have access to more reliable information for pricing bonds, assessing risk, and making informed investment decisions. Moreover, an improved understanding of the yield curve dynamics will aid in the development of financial products, trading strategies, and overall market

\section{Methodology}

% \subsection{Data imputation techniques}
% Parler un peu de methodes existante

% Limites de ces methods quand on veut calculer taux zero coupon

Numerous methods have been devised in the literature to calculate the yield curve. Nevertheless, when confronted with the task of computing the yield curve through the bootstrapping technique, a challenge arises in dealing with missing data. These missing data points pertain to maturities that were not issued or did not attract buyers, necessitating an imputation process.

Missing data mechanisms have implications on methods to handle the missing data \cite{nakagawa2015missing}. Mainly, we have three types of missingness mechanisms: 

\begin{itemize}
    \item Missing Completely at Random (MCAR): where missingness of a variable is completely independent of itself and other variables. In other words, the probability of missing data on a variable is unrelated to any other measured variable and is unrelated to the variable with missing values itself.
    \item Missing Not at Random (MNAR): where missingness of a variable is related to itself. 
    \item Missing at Random (MAR): where missingness of a variable is dependent on another variable. The probability of missing data on a variable is related to some other measured variable in the model, but not to the value of the variable with missing values itself.
\end{itemize}

For Burundi financial market, the missingness of some maturities is due to mainly when there is no one to buy the bonds, or if there is no issuance which if often influenced by the treasury needs. Thus, to encompass those data missingness scenarios, we considered the missingness data, to be missing completelty at random.

To handle missing data, various imputation methods from the literature are explored \cite{garcia2009k, josse2012handling, malarvizhi2012k, schmitt2015comparison}. The K nearest neighbour (KNN) approach involves selecting the K nearest observations (neighbors) based on a distance metric when dealing with an incomplete pattern. These chosen observations have known values for the features that need to be imputed. To estimate each incomplete feature value, a weighted average of these known values is calculated. Also the model  does not involve the creation of explicit predictive models as it relies on the training data-set itself as a "lazy" model. The methods available for handling missing values in principal component analysis (PCA) primarily offer point estimates of the parameters, such as axes and components, as well as estimates of the missing values. These methods provide single-value approximations rather than providing a range or distribution of potential values for the parameters or missing values. I this work, the following methods are considered for filling in the missing values:

\begin{itemize}
    \item Fill in missing values with previous or next value (with \textit{fill} function of \textbf{tidyr} package in R)
    
    \item Fill in missing values with the values of the nearest neighbours (with \textit{knnImputation}  function of \textbf{DMwR} package in R)

    \item Fill in missing values with Multivariate Imputation by Chained Equations, known as MICE \cite{van2011mice} (with \textit{mice}  function of \textbf{mice} package in R)

    \item Non-parametric Missing Value Imputation using Random Forest (with \textit{missForest}  function of \textbf{missForest} package in R)

    \item Linear regression with same maturity data issued previously.
    
\end{itemize}

To test these methods, a data-set of financial transactions with complete maturities is utilized. A random subset of the data is chosen to introduce missing maturities, based on the proportion of financial data with complete maturities from Burundi Central Bank, which stands at 65\% (which means, we had 35\% of missing data). Additionally, missing values for price and yield are generated. The data-set containing the created missing data is then employed to apply the six imputation methods. Following the imputation process, the accuracy of each method is assessed by calculating the Mean Absolute Error (MAE) between simulated and observed data. This computation is repeated multiple times to determine the distribution of error values for each method. Finally, the Shapiro test is conducted to examine whether the error values are normally distributed. If the $p-value > 0.05$, it implies that the distribution of the data are not significantly different from normal distribution. % Although it is known that small size sample may pass normality test is sensitive to sample size. Small samples most often pass normality tests. Therefore, it’s important to combine visual inspection and significance test in order to take the right decision.

% Shapiro-Wilk test is a hypothesis test that evaluates whether a data set is normally distributed. It evaluates data from a sample with the null hypothesis that the data set is normally distributed. A large p-value indicates the data set is normally distributed, a low p-value indicates that it isn’t normally distributed.

% In this  paper (Tackling Imputation Across Time Series Models
% Using Deep Learning and Ensemble Learning, cfr vos mails) the aUthors used a Aggregated Average Error (AGE) which is the simple average of the Root Mean Squared Error (RMSE) and the Mean Aboslute Error (MAE). Think about the use of the same tool (AGE) to gauge the accuracy of each method.

% Modele lineaire utilier: assumptions

% Estimer par les formules: les prix, redements, coupons avec les formules

% Estimer par les modeles lineaires: les prix, redements, coupons avec les modeles lineaire

\section{Results}

The following presents the results, showcasing the error values obtained from various models. These results enable us to assess the performance of each model concerning data imputation for the Burundian bond market.

% Results: les prix, redements, coupons avec les formules

% Results: les prix, redements, coupons avec les modeles lineaire

\begin{figure}[H] 
\centering
   \includegraphics[width=0.9\textwidth]{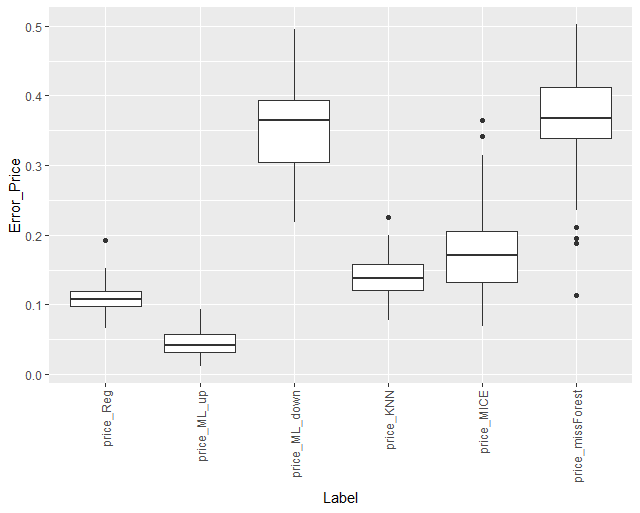}
   %\caption{}
   \label{fig:price}
\caption{Box-plots of the mean absolute error values computed for bonds prices after data imputation with six methods.}
\end{figure}

Based on Figure \ref{fig:price}, it appears that utilizing the previous values to fill the missing data results in low error values for the bond prices. This approach is closely followed by the implementation of linear regression.

%\newpage

\begin{figure}[H] 
\centering
\includegraphics[width=0.9\linewidth]{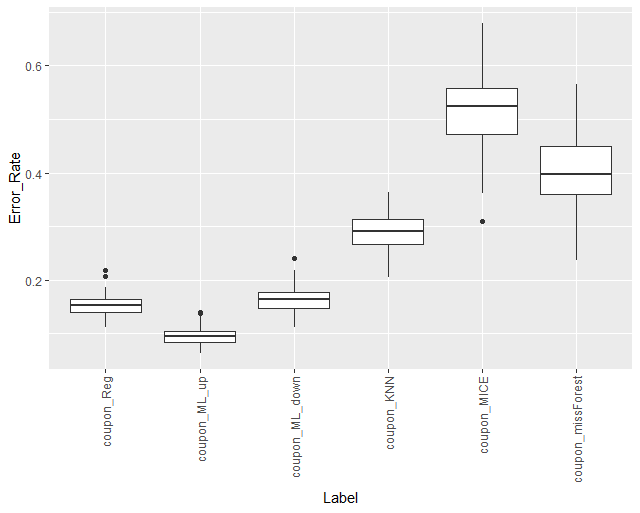}
\caption{Box plots of Mean absolute Error values computed for coupon rates after data imputation with six methods} 
\label{fig:rate}
\end{figure}

According to Figure \ref{fig:rate}, the error values for the bond prices show a similar trend of being low when employing the previous values for filling the missing data. Linear regression follows this method closely.

%\newpage

\begin{figure}[H] 
\centering
    \includegraphics[width=0.9\linewidth]{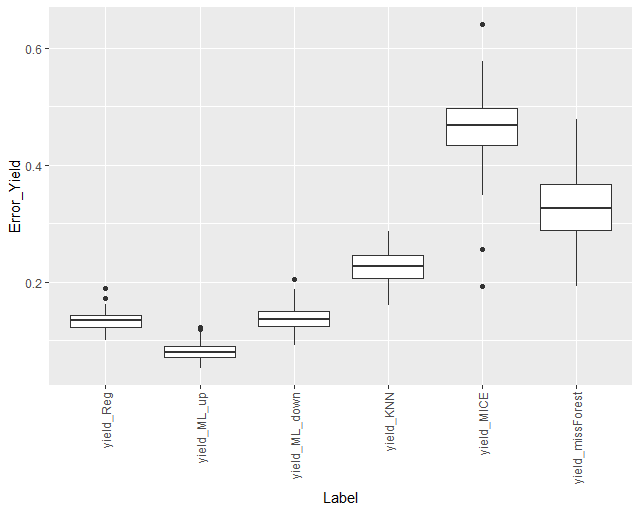}
\caption{Box plots of Mean absolute Error values computed for the yield after data imputation with six methods} 
\label{fig:yield}
\end{figure}

Based on Figure \ref{fig:yield}, it can be observed that utilizing the previous values to fill the missing data leads to low error values for the bond prices. This approach is closely followed by the implementation of linear regression.

%\newpage

In all three cases, employing the method of filling the missing data with previous values demonstrated superior performance compared to other methods. Table \ref{tab:error} illustrates that, overall, the error values for bond prices, yields, and coupon rates exhibited a normal distribution.

\begin{table}[H] 
\centering
\caption{Normality test for mean absolute error values computed after data imputation}

% {\small
      % \resizebox{\textwidth}{!}{
    \begin{tabular}{ p{5cm} p{4cm} p{3cm}}
    \hline
    \textbf{Method}  & \textbf{Estimate} & \textbf{p-value} \\ \hline

\multirow{3}{*}{Linear Regression} & Coupon & 0.7794  \\
 & Price & 0.0666 \\
 & Yield &  0.8639 \\

 \hline
\multirow{3}{*}{missForest} & Coupon & 0.8991  \\
 & Price & 0.0009 \\
 & Yield &  0.5212 \\
 
 \hline
\multirow{3}{*}{MICE} & Coupon & 0.4628  \\
 & Price & 0.0171\\
 & Yield &  0.9425 \\
\hline

\multirow{3}{*}{KNN} & Coupon & 0.6632  \\
 & Price & 0.4116 \\
 & Yield &  0.1631 \\
\hline

\multirow{3}{*}{Previous value} & Coupon & 0.9530  \\
 & Price & 0.0053 \\
 & Yield &  0.5682 \\
\hline

 \multirow{3}{*}{Next value} & Coupon & 0.0880 \\
 & Price & 0.8073 \\
 & Yield &  0.1118 \\

  \hline % OK

    \end{tabular} % } % }
    \label{tab:error}

\end{table}

% \newpage

\section{Discussion and conclusion}

Based on the results of the normality tests for mean absolute error (MAE) values computed after data imputation, we can compare the different methods in terms of how well they approximate a normal distribution for the error values. A higher p-value (greater than 0.05) indicates better adherence to the normality assumption. We compare the methods for each variable in the following subsections.

\subsection{Coupon rates} % Zero-coupon rates

The results of coupon rates show that for

\begin{itemize}
\item Linear Regression (p-value: 0.7794), the error values for coupon rates are likely to follow a normal distribution.
\item missForest (p-value: 0.8991), the error values for coupon rates are highly likely to follow a normal distribution.
\item MICE (p-value: 0.4628), the error values for coupon rates are likely to follow a normal distribution.
\item KNN (p-value: 0.6632), the error values for coupon rates are likely to follow a normal distribution.
\item Previous value (p-value: 0.9530), the error values for coupon rates are highly likely to follow a normal distribution.
\item Next value (p-value: 0.0880): The error values for coupon rates are not following a normal distribution.
\end{itemize}

For the coupon rate, the previous value method performs the best in terms of adhering to the normality assumption for the error values in the coupon rate variable. The missForest method also shows good results.

\subsection{Price}

The results of price of bonds show that for

\begin{itemize}
\item Linear Regression (p-value: 0.0666), the error values for bond prices are approximately normally distributed.
\item missForest (p-value: 0.0009), the error values for bond prices do not follow a normal distribution.
\item MICE (p-value: 0.0171), the error values for bond prices are approximately normally distributed.
\item KNN (p-value: 0.4116), the error values for bond prices are likely to follow a normal distribution.
\item Previous value (p-value: 0.0053), the error values for bond prices are approximately normally distributed.
\item Next value (p-value: 0.8073), the error values for bond prices are likely to follow a normal distribution.
\end{itemize}

For the bond price, we can see that the Linear Regression method and the Previous value method both perform well in terms of normality for the error values in the price variable. The missForest method, on the other hand, shows the poorest adherence to the normality assumption for this variable.

\subsection{Yield}

The results of yield from the bonds show that for

\begin{itemize}
\item  Linear Regression (p-value: 0.8639), the error values for yields are highly likely to follow a normal distribution.
\item missForest (p-value: 0.5212), the error values for yields are likely to follow a normal distribution.
\item MICE (p-value: 0.9425), the error values for yields are highly likely to follow a normal distribution.
\item KNN (p-value: 0.1631), the error values for yields are likely to follow a normal distribution.
\item Previous value (p-value: 0.5682), the error values for yields are likely to follow a normal distribution.
\item Next value (p-value: 0.1118), the error values for yields are not following a normal distribution.
\end{itemize}

For the yield curve, we realize that the Linear Regression method and the MICE method both show strong results in adhering to the normality assumption for the error values in the yield variable. The KNN method and the Previous value method also exhibit reasonably good performance.

Overall, the Linear Regression method and the Previous value method consistently show better results across the variables, with higher p-values indicating better adherence to the normality assumption for the error values. The missForest method performs well for coupon rates but poorly for bond prices. The Next value method shows mixed results, with better performance for bond prices and poorer performance for yields.

These results argue the findings of \cite{schmitt2015comparison}, which indicate that the most popular methods (KNN, MICE) are not necessarily the most efficient ones even thouth they did not compare it with the previous value and the linear regression method.

In conclusion, the normality tests indicate that both the Linear Regression method and the Previous value method are recommended for imputing missing data in this study. These methods consistently generate error values that closely approximate a normal distribution across the variables. However, the Linear Regression method stands out as the preferred choice, not only because of its ability to approximate normality but also due to its predictive capabilities in estimating the missing data. Researchers should consider these results when selecting an appropriate imputation method for similar financial market data analyses.

\bibliographystyle{elsarticle-num-names}
\bibliography{missing_data_financial_market.bib}
%\input{appendix}





% \end{thebibliography}

\end{document}